\documentclass[preprint,aps,prl,showpacs,amsfonts,amssymb,superscriptaddress]{revtex4}
\usepackage{graphicx}

\begin{document}
\title{Electrically injected cavity polaritons}
\author{L. Sapienza}
\affiliation{Laboratoire ``Mat\'eriaux et Ph\'enom\`enes
Quantiques'', Universit\'e Paris Diderot-Paris 7, 75013 Paris,
France} \affiliation{CNRS, UMR 7162, 75013 Paris, France}
\author{A. Vasanelli}
\email{Angela.Vasanelli@univ-paris-diderot.fr}
\affiliation{Laboratoire ``Mat\'eriaux et Ph\'enom\`enes
Quantiques'', Universit\'e Paris Diderot-Paris 7, 75013 Paris,
France} \affiliation{CNRS, UMR 7162, 75013 Paris, France}
\author{R. Colombelli}
\affiliation{Institut d'Electronique Fondamentale, Universit\'e
Paris Sud, CNRS, 91405 Orsay, France}
\author{C. Ciuti}
\affiliation{Laboratoire ``Mat\'eriaux et Ph\'enom\`enes
Quantiques'', Universit\'e Paris Diderot-Paris 7, 75013 Paris,
France} \affiliation{CNRS, UMR 7162, 75013 Paris, France}
\author{Y. Chassagneux}
\affiliation{Institut d'Electronique Fondamentale, Universit\'e
Paris Sud, CNRS, 91405 Orsay, France}
\author{C. Manquest}
\affiliation{Laboratoire ``Mat\'eriaux et Ph\'enom\`enes
Quantiques'', Universit\'e Paris Diderot-Paris 7, 75013 Paris,
France} \affiliation{CNRS, UMR 7162, 75013 Paris, France}
\author{U. Gennser}
\affiliation{Laboratoire de Photonique et Nanostructures,
LPN-CNRS, Route de Nozay, 91460 Marcoussis, France}
\author{C. Sirtori}
\affiliation{Laboratoire ``Mat\'eriaux et Ph\'enom\`enes
Quantiques'', Universit\'e Paris Diderot-Paris 7, 75013 Paris,
France} \affiliation{CNRS, UMR 7162, 75013 Paris, France}

\date{\today}

\begin{abstract}
We have realised a semiconductor quantum structure that produces
electroluminescence while operating in the light-matter strong
coupling regime. The mid-infrared light emitting device is
composed of a quantum cascade structure embedded in a planar
microcavity, based on the GaAs/AlGaAs material system. At zero
bias, the structure is characterised using reflectivity
measurements which show, up to room temperature, a wide polariton
anticrossing between an intersubband transition and the resonant
cavity photon mode. Under electrical injection the spectral
features of the emitted light change drastically, as electrons are
resonantly injected in a reduced part of the polariton branches.
Our experiment demonstrates that electrons can be selectively
injected into polariton states up to room temperature.
\end{abstract}

\maketitle The study of electron - hole excitations strongly
coupled to photonic modes in a semiconductor micro-cavity, has
motivated a wealth of fascinating
experiments~\cite{Weisbuch,savvidis,bose} in the past ten years.
Recently, the light - matter strong interaction has been observed
also in a two-dimensional electron gas (2DEG) coupled to a cavity
photon mode~\cite{dini}. These excitations, called intersubband
(ISB) polaritons~\cite{ciuti_PRB}, have been experimentally
demonstrated using angle-dependent reflectance spectroscopy of
multiple 2DEGs~\cite{anappara_APL2005,anappara_APL2006}. Recently,
a photovoltaic probe of intersubband polaritons in a quantum
cascade structure has also been realized~\cite{sapienza}.

The strong coupling regime may be especially interesting for light
emitting devices (LED) based on ISB transitions. Their radiative
efficiency is in fact very poor ($\approx 10^{-5}$) because
non-radiative phenomena control the lifetime of electrons in
excited subbands (the radiative lifetime is $\approx 10-100$~ns in
the mid-infrared while the non-radiative one is $\approx 1$~ps).
In the strong-coupling regime, the light-matter interaction can be
dominated by the time of a Rabi oscillation rather than the
radiative lifetime, with a possible advantage in term of external
efficiency. The possibility of a stimulated emission of ISB
polaritons could also lead to inversionless mid and far infrared
lasers, with lower thresholds with respect to quantum cascade
lasers. In the context of light sources, devices based on
electrical injection, rather than optical excitation, would be
more suitable for applications.

Up to now, electrical injection of a micro-cavity LED in the
strong coupling regime has only been reported in organic
semiconductors~\cite{nurmikko}. Moreover, the description of a
current injecting electrons into a polaritonic system constitutes
a new physical framework that lacks a complete theoretical
description. A first attempt to describe the coupling between an
electronic reservoir and intersubband polaritons has been proposed
in ref.~\cite{ciuti_PRA}. The authors derive an analytical
expression for the electroluminescence by introducing the coupling
of the polaritonic system with a bath of electronic excitations.

In the present paper, we report on the experimental realization of
a GaAs/AlGaAs electroluminescent device working in the
light-matter strong coupling regime. A quantum cascade (QC)
structure is used as an active region, as proposed in
ref.~\cite{colombelli}. Clear differences between
electroluminescence (EL) and absorption spectra of the same
structure are observed, thus pointing out the role of the
electrical injection. A very good agreement between the
experimental results and the simulations is found with a
phenomenological model, in which the EL spectra are obtained as
the product of the absorption spectra with a Gaussian filter
function.

\begin{figure}[ht]
\includegraphics[width=0.95\columnwidth]{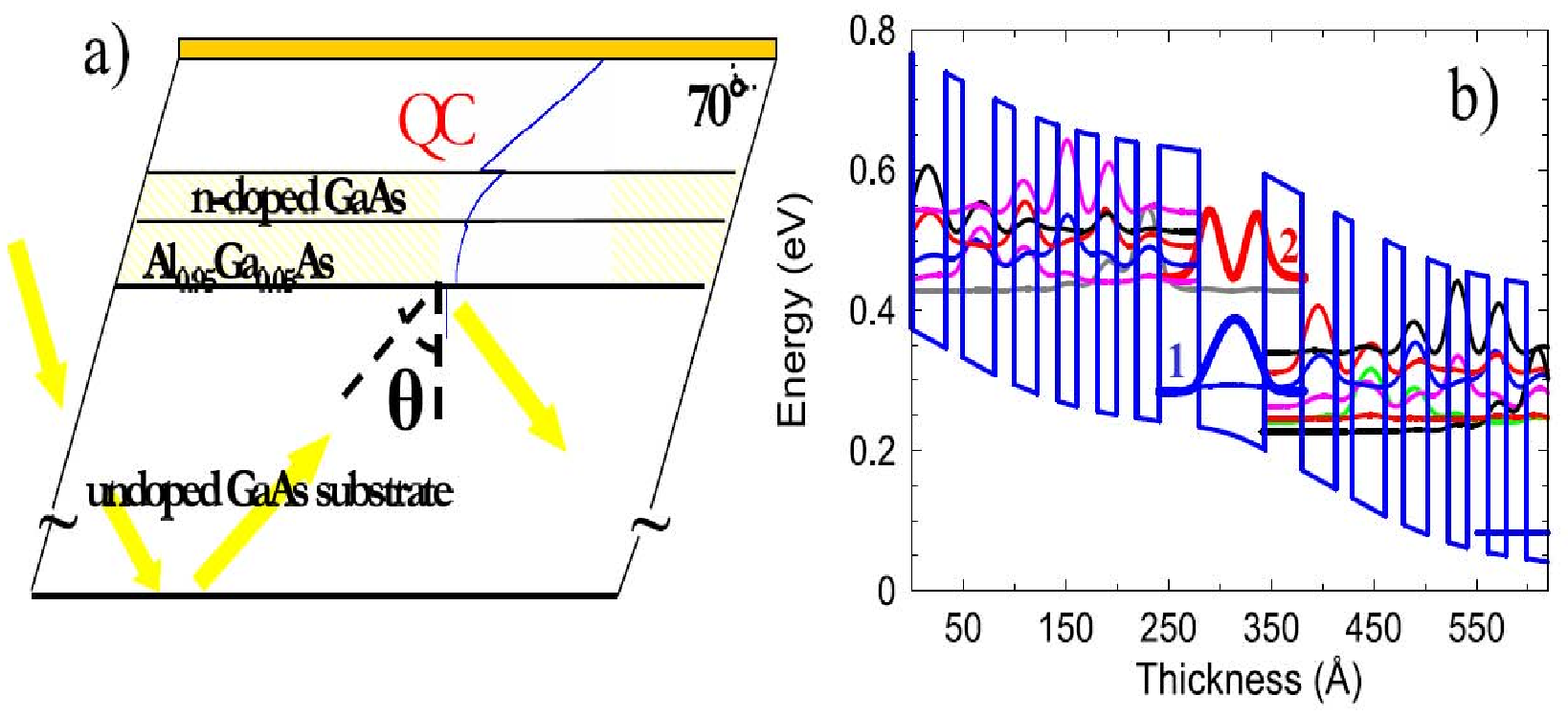}
\caption{(a) Schematic view of the sample. The arrows represent
the optical path of the incident beam in the reflectivity
measurements. The top and the bottom mirror are shaded. The
thickness of the $n$-doped (3x10$^{18}$~cm$^{-3}$) and of the
Al$_{0.95}$Ga$_{0.05}$As layers are 0.56~$\mu$m and 0.52~$\mu$m,
respectively. A 86~nm layer of 1x10$^{17}$~cm$^{-3}$ Si-doped GaAs
and a 17~nm layer of 3x10$^{18}$~cm$^{-3}$ Si-doped GaAs are grown
on top of the QC-LED to allow current flow towards the metallic
contacts. The continuous (blue) line represents the intensity of
the photon mode along the growth axis. (b) Band diagram of the
QC-LED for an applied voltage of 6~V. The square moduli of the
relevant wavefunctions are shown. The layer sequence of one period
of the structure, in nm, from left to right, starting from the
largest well is
6.4/\textbf{3.6}/3.3/\textbf{1.6}/\underline{3.2}/\underline{\textbf{1.8}}/\underline{2.3}/\underline{\textbf{2.0}}/\underline{1.9}/\underline{\textbf{2.0}}/1.8/\textbf{2.0}/2.2/\textbf{3.9}.
Al$_{0.45}$Ga$_{0.55}$As layers are in bold, underlined layers are
$n$ doped with Si $N_d = 3 \times 10^{17}$~cm$^{-3}$. This
sequence has been repeated 30 times in the sample.}
\label{fig_struc}
\end{figure}

The sample (sketched in fig.~\ref{fig_struc}a) is composed of a QC
LED grown by molecular-beam epitaxy, on an undoped GaAs (001)
substrate. The QC LED is placed within a metal/semiconductor
planar micro-cavity, designed for the confinement of transverse
magnetic polarized radiation. Light confinement is realized by
sandwiching the QC structure between a bottom reflector made of a
low refractive index Si-doped GaAs and Al$_{0.95}$Ga$_{0.05}$As
layers and a top metallic mirror evaporated on the surface. The
latter is also used as top electrical contact.

The band diagram of the QC LED, obtained with self-consistent
Schr\"{o}dinger-Poisson calculations, is presented in
fig.~\ref{fig_struc}b under an applied bias of 6~V. The radiative
transition takes place in the largest well between the states
labelled 1 and 2, whose nominal energy separation is
$E_{12}=160$~meV. Electrical injection into the subband 2 is
obtained by resonant tunnelling through an injection
barrier~\cite{sirtori_IEEE}. Electrons are extracted from subband
1 through the states of a miniband, which also has the purpose to
inject electrons into the following period. The injection region
has been designed in order to have a tunnelling time out of the
subband 1 longer than the scattering time from 2 to 1 in order to
avoid population inversion and accumulate electrons into the
subband 1. This is an essential feature for the observation of the
strong coupling regime~\cite{colombelli}. Furthermore, the
injection region has been highly doped ($7 \times
10^{11}$~cm$^{-2}$, see caption of fig.~\ref{fig_struc}) in order
to provide a high density 2DEG in the well. As a consequence, the
sample is also suitable for reflectivity measurements when no bias
is applied~\cite{nota_bd}.

A photon confined within the micro-cavity can be absorbed
promoting an electron to level 2, with the creation of an ISB
excitation on top of the 2DEG. By varying the propagating angle of
the light within the cavity, the energy of the photonic mode can
be tuned across the energy of the ISB excitation. If the strong
coupling regime is achieved, the degeneracy of the excitation and
photonic states at resonance will be removed and the two branches
will anticross. Angle resolved measurements will thus reveal the
presence of intersubband polaritons in the system.

\begin{figure}[ht]
\includegraphics[width=0.95\columnwidth]{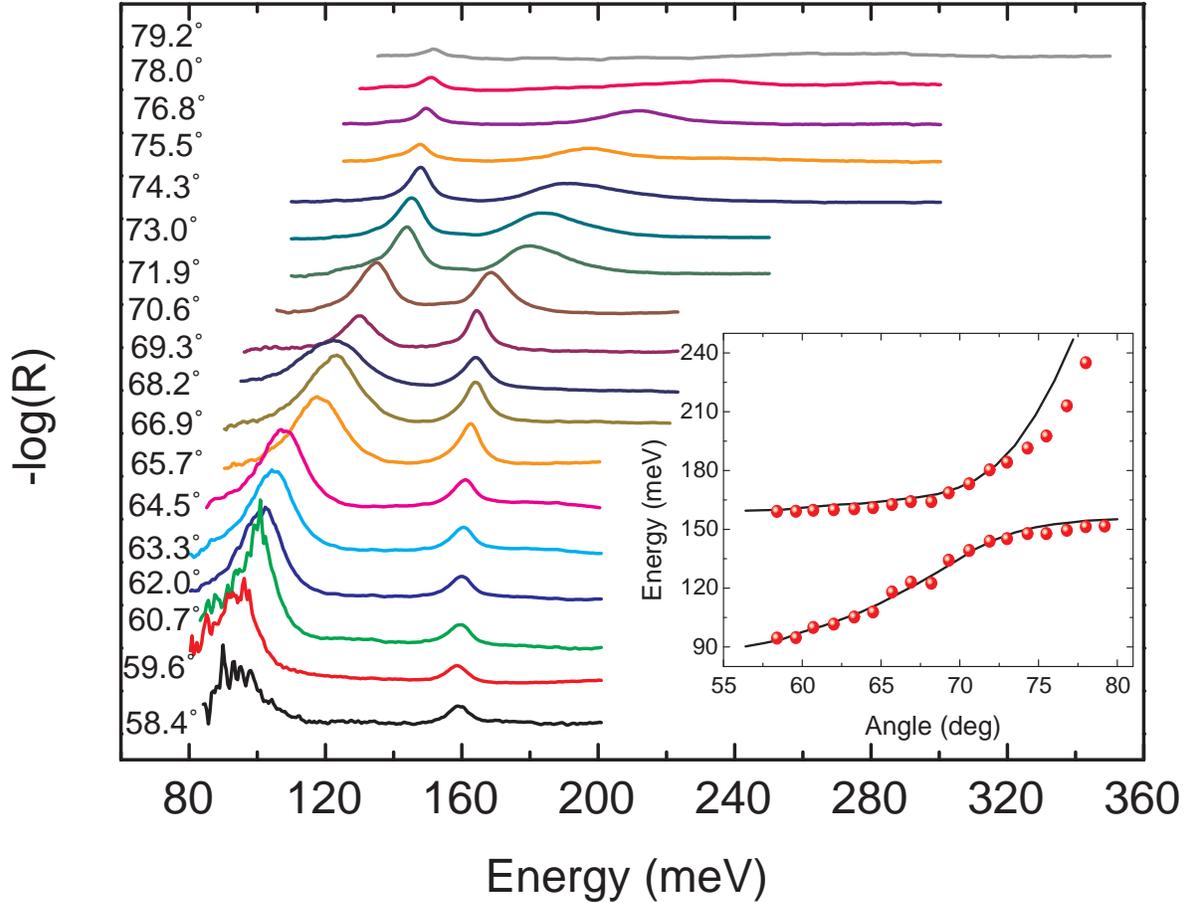}
\caption{Reflectivity spectra as a function of the incident angle,
at 78~K. The spectra result from the ratio $R$ between spectra
collected with TM and TE-polarized incident light and are offset
each other for clarity. A Nicolet Fourier-transform infrared
spectrometer, in rapid scan mode, with a spectral resolution of
8~cm$^{-1}$, has been used. The radiation of a Globar lamp is
focused on the facet of the sample with a $f/1.5$ ZnSe lens and
the transmitted light is collected with a $f/1.0$ ZnSe lens. In
the inset, the energy position of the reflectivity peaks is
plotted as a function of the angle (full dots), as well as the
results of transfer matrix calculations (line).} \label{abs}
\end{figure}

To this end, the sample has been prepared with $70 ^\circ$
polished facets (see fig.~\ref{fig_struc}a) and a metallic layer
[Ti(10~nm)/Au(500~nm)] was e-beam evaporated on the top surface,
for standard reflectivity measurements~\cite{liu}. Angle resolved
reflectivity spectra, performed at 78~K, are shown in
fig.~\ref{abs}. Two peaks are clearly visible across a very wide
angular range (almost the whole range where the cavity operates,
between total internal reflection at $\approx 59 ^\circ$ and
Brewster angle at $\approx 87 ^\circ$). In the inset of
fig.~\ref{abs} we have plotted the energy position of the two
peaks (symbols) as a function of the internal angle in the cavity,
showing an anticrossing feature at approximately
$71^\circ$°~\cite{footnote}.
We
attribute the two branches in the inset of fig.~\ref{abs} to the
upper and lower polaritons (UP and LP respectively). It is worth
mentioning that the strong coupling regime has also been observed
in reflectivity up to 300~K with very similar features to those
shown in fig.~\ref{abs}. In the inset of fig.~\ref{abs} we have
also plotted the LP and UP branches calculated in the transfer
matrix formalism (line). In the simulations, the contribution of
the ISB transition has been taken into account in the dielectric
permittivity of the quantum well layers including an additional
term in the form of an ensemble of classical polarized Lorentz
oscillators, as in ref.~\cite{dini}. The dispersions of the
Au~\cite{ordal} and of the doped layers~\cite{Palik} have also
been included. In the calculations we used an electronic density
in the fundamental subband $N_{2DEG}=6 \times 10^{11}$~cm$^{-2}$
to reproduce the experimental results. This value is consistent
with the results of Shubnikov-de Haas measurements which give 7.2
$\times 10^{11}$~cm$^{-2}$.

The energy splitting deduced from the inset of fig.~\ref{abs} is
34~meV. It is worth noticing that this value {\em{is not}} the
vacuum field Rabi splitting $2 \hbar \Omega_R$, which has to be
deduced from the $E \, \, vs \, \, k$ dispersion relation, where
$k$ is the in-plane momentum of the photon. The relationship
between $k$ and the light propagating angle inside the cavity
$\theta$ reads: $k=n \, E \sin \theta /(c
\hbar)$~\cite{ciuti_PRB}, where $n$ is the cavity refractive index
and $E$ the energy of the considered excitation branch. This
relationship introduces a strong distortion of the polariton
branches, resulting in a value of the Rabi splitting $2 \hbar
\Omega_R = 11$~meV.

\begin{figure}[ht]
\includegraphics[width=0.85\columnwidth]{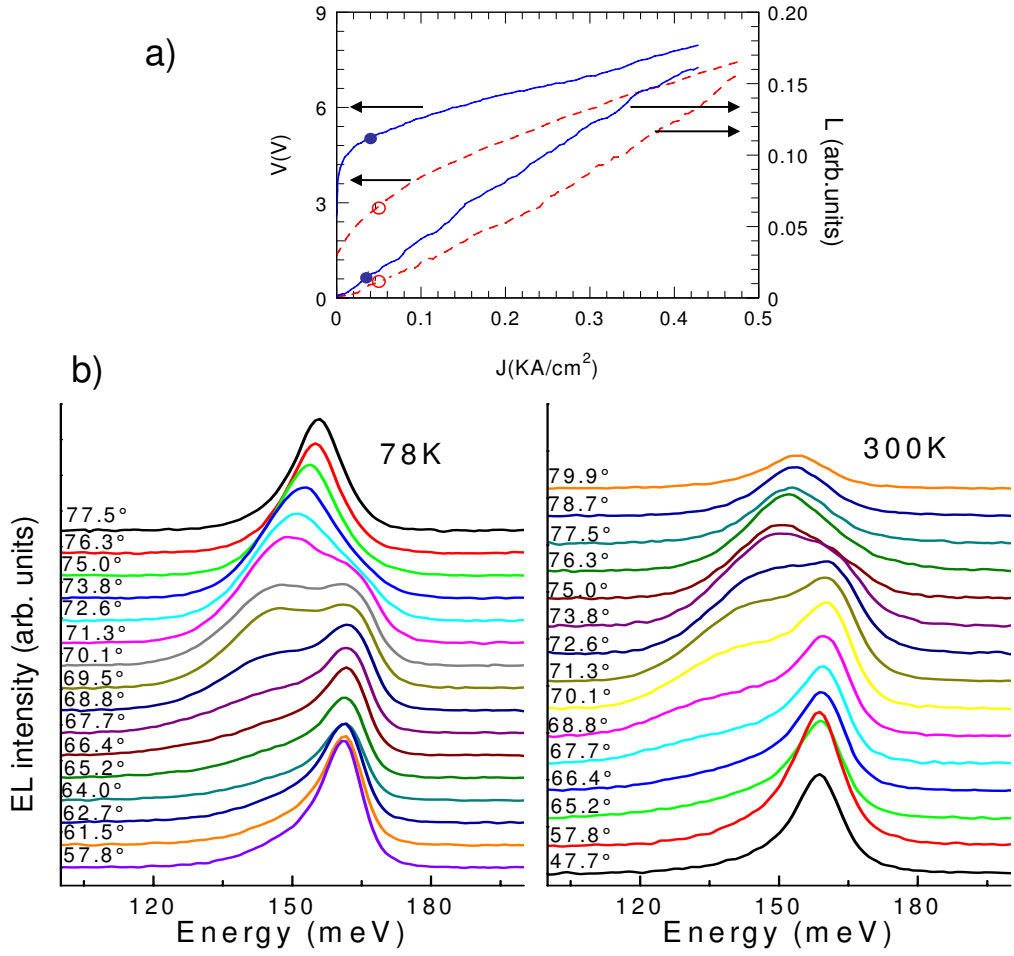}
\caption{a) Voltage and electroluminescence intensity as a
function of the injected current density in pulsed operation (20
kHz, 500 ns) at 73.8$^\circ$ at 78~K (continuous line) and 300~K
(dashed line). The dots indicate the working conditions for
obtaining the electroluminescence spectra shown below.
b)Electroluminescence spectra, offset each other for clarity, as
functions of the propagating angle within the micro-cavity, at 78K
and 5~V (left side) and at 300~K for an applied voltage of 2.8~V
(right side). These voltages have been chosen as the lowest
possible values to measure a spectrum at the two temperatures. In
order to increase the electroluminescence signal, we worked at ~50
\% duty cycle (100~kHz repetition rate and 5~$\mu$s pulse width).
Spectra are collected in step scan mode, with a spectral
resolution of 16~cm$^{-1}$; a $f/1.5$ ZnSe lens is used as
collecting lens. } \label{emissione}
\end{figure}

By applying an external bias to the device, the intersubband
strong coupling regime can be investigated in emission, via EL
measurements. The sample is thus mesa etched (circular 220~$\mu$m
diameter mesas), metallic contacts
[Ni(10~nm)/Ge(60~nm)/Au(120~nm)/Ni(20~nm)/Au(200~nm)] are
evaporated on top to allow current flow and the facet is polished
with an angle of $70^\circ$. The sample is soldered onto a copper
holder, mounted onto the cold finger of a cryostat and the
temperature is varied from 10~K to 300~K. The electrical and
optical characteristics of the device at 78~K and 300~K are
summarized in fig.~\ref{emissione}a, where the evolution of the
voltage and of the electroluminescence intensity as functions of
the injected current density is shown. The structure presents a
diode-like behavior, with an alignment at 5~V at 78~K. The output
power shows a well defined linear dependence with the injected
current. The EL spectra have been obtained by applying to the
structure a voltage close to the alignment voltage (the working
conditions for obtaining the spectra are shown by dots in
fig.~\ref{emissione}a), in order to keep the electronic density in
the lowest subband as high as possible.

Fig.~\ref{emissione}b presents the EL spectra measured at 78~K for
an applied voltage of 5~V (left side) and at 300~K for an applied
voltage of 2.8~V (right side) at different angles. The spectral
features show a very strong angular dependence. For the angles
close to $70^ \circ$ we can observe at both temperatures two
peaks, which are quite well separated in energy. On the contrary,
when moving away from the resonance, only one peak is observed
with a large shoulder.

\begin{figure}[ht]
\includegraphics[width=0.85\columnwidth]{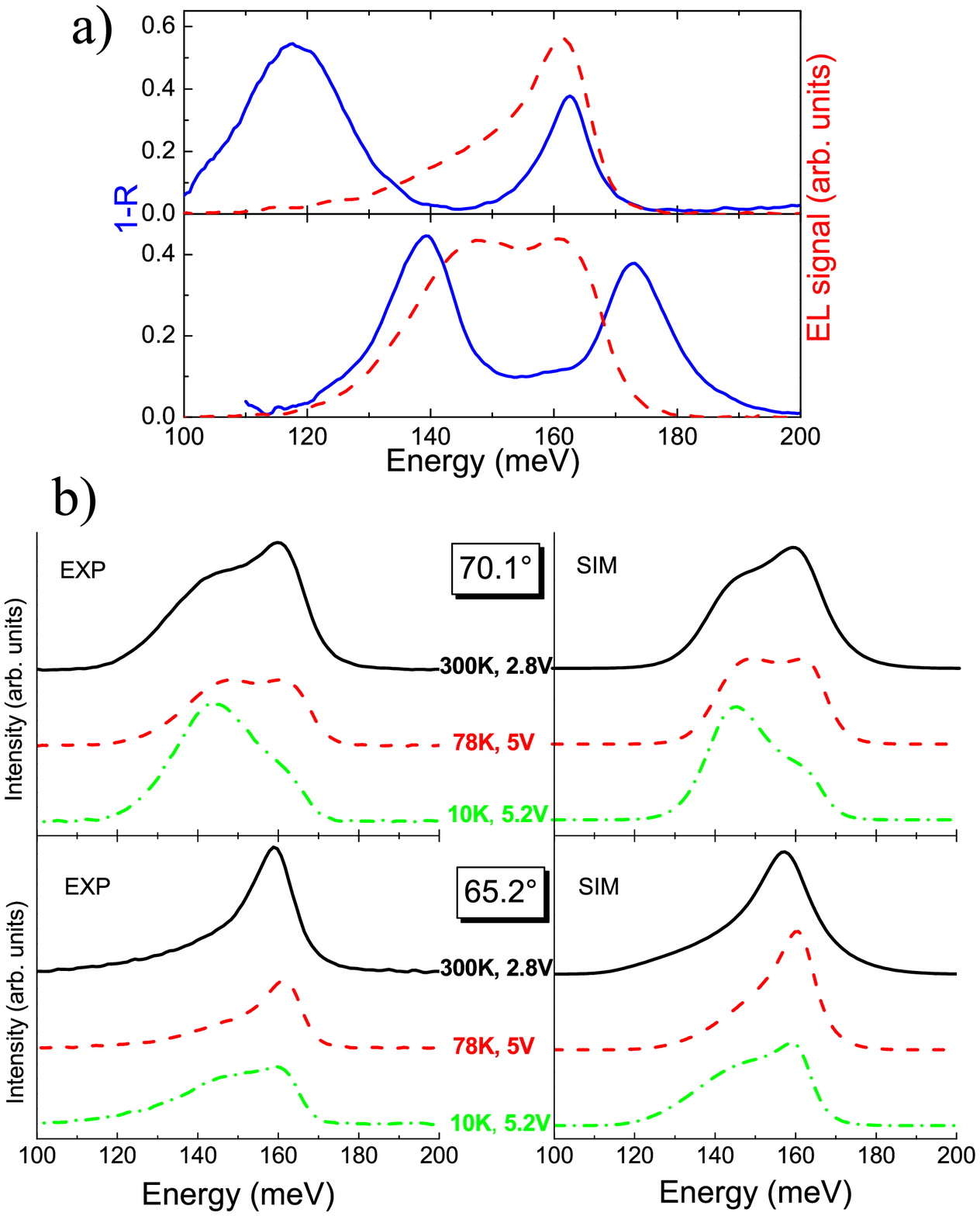}
\caption{a) Comparison between $1-R$ (i.e. absorption, see text)
and electroluminescence spectra measured at 78~K, at ~$65^\circ$
(upper panel, reflectivity $65.7^\circ$; electroluminescence
$65.2^\circ$) and at ~$70^\circ$ (lower panel: reflectivity
$70.6^\circ$, electroluminescence $70.1^\circ$). b) Experimental
(left side) and simulated (right side) electroluminescence.}
\label{sim}
\end{figure}

The EL spectra here presented are substantially different from the
results obtained in reflectivity: it is important to underline
that in reflectivity the polaritons are visible from about 90 to
240~meV, while in EL spectra the energy range where the
experimental features are observable is approximately restricted
within $120$~meV and $170$~meV. Furthermore, this energy interval
(or "energy window") is different in the two sets of data shown in
fig.~\ref{emissione} and depends on the voltage applied to the
structure and on the temperature. In order to highlight this
point, we show in fig.~\ref{sim}a a comparison between
reflectivity and EL spectra measured at 78~K. In the top panel,
the spectra at $65\,^{\circ}$ are presented. In reflectivity, we
observe the UP and LP peaks, while the electroluminescence
spectrum shows a single peak close to the energy of the upper
polariton with a shoulder towards lower energies. In the lower
panel, spectra at $70.1\,^{\circ}$ are presented: we observe the
presence of two peaks in electroluminescence, with an energy
difference of about 13~meV, a value much lower than the energy
difference between the reflectivity peaks (34~meV).

The substantial difference between EL and reflectivity
measurements originate from the coupling that the electronic
component of the polariton states has with the injection region.
In fact, for a fixed voltage, only one of the states of the
injection region has a relevant probability to tunnel through the
injection barrier, thus populating the polariton states. In order
to simulate the EL spectra, we describe the injector as a Gaussian
function $G(E)=\exp{\left(-(E-E_0)^2/(2\sigma^2)\right)}$, where
$E_0$ is related to the energy position of the injector state with
respect to the subband 1; $\sigma$ reflects the broadening of the
states in the injector due to disorder arising from interface and
impurity scattering~\cite{faist_APL}. After simulating the
absorption spectrum by using the transfer matrix formalism
(obtained as $1-R$, by neglecting the transmission through the
upper mirror), we calculate the EL spectrum as the product between
the Gaussian function and the absorption spectrum:
$S(E)=G(E)(1-R)$. The parameters of the Gaussian function, $E_0$
and $\sigma$, are kept as fit parameters and determined by the
comparison between the simulated and the measured EL spectra. We
show in fig.~\ref{sim}b the measured (left side) and simulated
(right side) electroluminescence, for three different temperatures
(and voltages), at the same angles as in fig.~\ref{sim}a. We
reproduce very well the shapes of the measured spectra in the
entire angular range in the three temperature and voltage regimes.
The energy position of $G(E)$ has been fixed respectively to
$E_0=150$~meV (10~K, 5.2~V), $E_0=155$~meV (78~K, 5~V),
$E_0=158$~meV (300~K, 2.8~V). In the three cases, $E_0$ is close
to the energy of the ISB transition ($E_{12}=160$~meV at 10~K and
78~K, $E_{12}=156$~meV at 300~K). The reason for the differences
in the values of $E_0$ is due to a different alignment of the
states of the injector determined by the different applied bias. A
good agreement with the experimental data is found with
$\sigma=10$~meV for 10~K and 78~K, and $\sigma=14$~meV for 300~K.
In the simulations, we used an electronic density reduced by a
factor of 3 ($N_{2DEG}=2 \times 10^{11}$~cm$^{-2}$) as compared to
the unbiased system; this value has been determined by a
calculation of the electronic population as a function of the
electric field applied to the structure~\cite{aude} and by a
comparison between the simulated and measured electroluminescence
spectra.

The phenomenological model described above can also be understood
in the input-output theoretical framework developed in
ref.~\cite{ciuti_PRA}. An analytical relationship between the EL
and the absorption spectra may be found by exploiting the
unitarity property of the input-output matrix~\cite{ciuti_PRA} to
describe the coupling of the system with an electronic reservoir.
In this formalism it can be shown that the EL spectrum takes the
form: ${\mathcal L}(E) \propto {\mathcal A}(E) I^{el}_{exc}(E)$,
where ${\mathcal A}(\omega)$ is the microcavity absorption
spectrum and $I^{el}_{exc}(\omega)$ is related to the spectral
properties of the electronic reservoir. This expression is
analogue to the one used to interpret our experimental
results~\cite{footnote2}.

The very good agreement between the simulated and the measured
spectra of fig.~\ref{sim} is a strong proof of the observation of
polaritonic luminescence, based on an electrical injection, up to
room temperature. Furthermore, our experiment demonstrates that
electrons can be selectively injected into polariton states.

In conclusion, we have realized a semiconductor EL device based on
the light-matter strong coupling regime. The achievement of this
regime has been demonstrated by reflectivity measurements, up to
room temperature. Electroluminescence measurements show an
electrical injection into polariton states up to room temperature.
Our experimental results have been interpreted by using a
phenomenological model, in which an energy filter related to the
electrical injection is introduced. We believe that these results
open the way to a new class of devices operating in the strong
coupling regime. In the future, an optimized electronic resonant
injection into the polariton branches could pave the way to the
realization of an electrically pumped polariton laser.

\begin{acknowledgments}
The authors thank I.~Carusotto, H.~C.~Liu, I.~Sagnes, N.~Alayli,
L.~Largeau, O.~Mauguin, S.~De~Liberato for fruitful discussions
and help. The device fabrication has been performed at the
nano-center \textit{Centrale Technologique Minerve} at the
Institut d'Electronique Fondamentale. We gratefully acknowledge
support from EU MRTN-CT-2004-51240 POISE and ANR-05-NANO-049
INTERPOL.
\end{acknowledgments}

\newpage

\begin{thebibliography}{30}
\bibitem{Weisbuch} C.~Weisbuch, M.~Nishioka, A.~Ishikawa, and Y.~Arakawa, Phys. Rev. Lett. \textbf{69}, 3314
(1992).
\bibitem{savvidis} P.~G.~Savvidis, J.~J.~Baumberg,
R.~M.~Stevenson, M.~S.~Skolnick, D.~M.~Whittaker and
J.~S.~Roberts, Phys.~Rev.~Lett. {\textbf{84}}, 1547 (2000).
\bibitem{bose} J.~Kasprzak, M.~Richard, S.~Kundermann, A.~Baas, P.~Jeambrun, J.~M.~J.~Keeling, F.~M.~Marchetti,
M.~H.~Szyman$\acute{s}$ka, R.~Andr\'e, J.~L.~Staehli, V.~Savona,
P.~B.~Littlewood, B.~Deveaud and Le~Si~Dang, Nature \textbf{443},
409 (2006).
\bibitem{dini} D.~Dini, R.~K\"{o}hler, A.~Tredicucci, G.~Biasiol and L.~Sorba, Phys.~Rev.~Lett. \textbf{90}, 116401
(2003).
\bibitem{ciuti_PRB} C.~Ciuti, G.~Bastard and I.~Carusotto, Phys.~Rev. B \textbf{72}, 115303
(2005).
\bibitem{anappara_APL2005} A. A. Anappara, A.
Tredicucci, G. Biasiol and L. Sorba, Appl. Phys. Lett.
\textbf{87}, 051105 (2005).
\bibitem{anappara_APL2006} A. A. Anappara, A.
Tredicucci, F.~Beltram, G. Biasiol and L. Sorba, Appl. Phys. Lett.
\textbf{89}, 171109 (2006).
\bibitem{sapienza} L.~Sapienza, A.~Vasanelli, C.~Ciuti,
C.~Manquest, C.~Sirtori, R.~Colombelli and U.~Gennser,
Appl.~Phys.~Lett. \textbf{90}, 201101 (2007).
\bibitem{nurmikko} J.~R.~Tischler, M.~S.~Bradley, V.~Bulovic, J.~H.~Song and A.~Nurmikko, Phys. Rev. Lett. \textbf{95}, 036401
(2005).
\bibitem{ciuti_PRA} C.~Ciuti and I.~Carusotto, Phys.~Rev.~A
\textbf{74}, 033811 (2006).
\bibitem{colombelli} R. Colombelli, C. Ciuti, Y. Chassagneux and C. Sirtori, Semicond. Sci. Technol. \textbf{20}, 985
(2005).
\bibitem{sirtori_IEEE} C.~Sirtori, F.~Capasso, J.~Faist,
A.~L.~Hutchinson, D.~L.~Sivco and A.~Y.~Cho, IEEE Journal of
Quantum Electronics \textbf{34}, 1722 (1998).
\bibitem{nota_bd} For the band
diagram at zero bias see ref.~\cite{sapienza}.
\bibitem{liu} Intersubband Transitions in Quantum Wells: Physics and Device Applications I, edited by
H.C.Liu and F.Capasso, Semiconductors and Semimetals vol.62,
Academic Press, San Diego (2000).
\bibitem{footnote} The
mixing of photonic and material excitation components is also
evident from the behaviour of the linewidths as functions of the
angle: the UP and LP show the same linewidth at resonance, when
the two contributions are equally present in the polaritonic
states.
\bibitem{ordal} M.~A.~Ordal, R.~J.~Bell, R.~W.~Alexander,
L.~L.~Long and M.~Querry, Appl. Optics \textbf{26}, 744 (1987).
\bibitem{Palik} \textit{Handbook of Optical Constants of Solids}, edited by
E.~D.~Palik, Academic Press (1998).
\bibitem{faist_APL} J.~Faist, F.~Capasso, C.~Sirtori, D.~L.~Sivco, A.~L.~Hutchinson, S.~N.~G.~Chu, and
A.~Y.~Cho, Appl.~Phys.~Lett. \textbf{65}, 94 (1994).
\bibitem{aude} A. Leuliet, private communication.
\bibitem{footnote2} A more detailed theoretical description of the electrical
injection into polaritonic states is beyond the scope of the
present article and it is in progress.
\end{thebibliography}
\end{document}